\documentclass[aps,prl,twocolumn,showpacs]{revtex4}
\usepackage{graphicx}
\usepackage{dcolumn}

\begin{document}

\preprint{APS/123-QED}

\title{Anomalous lattice response at the Mott Transition in a Quasi-2D Organic Conductor
}

\author{M. de Souza$^{1}$}
\author{A. Br\"{u}hl$^{1}$}
\author{Ch. Strack$^{1}$}
\author{B. Wolf$^{1}$}
\author{D. Schweitzer$^{2}$}
\author{M. Lang$^{1}$}

\address{$^{1}$Physikalisches Institut, J.W. Goethe-Universit\"{a}t
Frankfurt(M), FOR 412, D-60054 Frankfurt am Main, Germany}
\address{$^{2}$3. Physikalisches Institut, Universit\"{a}t
Stuttgart, D-70550 Stuttgart, Germany}


\date{\today}

\begin{abstract}
Discontinuous changes of the lattice parameters at the Mott
metal-insulator transition are detected by high-resolution
dilatometry on deuterated crystals of the layered organic
conductor $\kappa$-(BEDT-TTF)$_{2}$Cu[N(CN)$_{2}$]Br. The uniaxial
expansivities uncover a striking and unexpected anisotropy,
notably a zero-effect along the in-plane $c$-axis along which the
electronic interactions are relatively strong. A huge thermal
expansion anomaly is observed near the end-point of the
first-order transition line enabling to explore the critical
behavior with very high sensitivity. The analysis yields critical
fluctuations with an exponent $\tilde{\alpha } \simeq$ 0.8 $\pm$
0.15 at odds with the novel criticality recently proposed for
these materials [Kagawa \textit{et al.}, Nature \textbf{436}, 534
(2005)]. Our data suggest an intricate role of the lattice degrees
of freedom in the Mott transition for the present materials.
\end{abstract}

\pacs{72.15.Eb, 72.80.-r, 72.80.Le, 74.70.Kn}

\maketitle

The Mott metal-insulator (MI) transition has been the subject of
intensive research for many years, see e.g.\,\cite{Imada 98} for a
review. Materials intensively discussed in this context include
transition metal oxides, notably Cr-doped V$_{2}$O$_{3}$, and,
recently, organic $\kappa$-(BEDT-TTF)$_{2}$X charge-transfer salts
\cite{Lefebvre 00,Limelette 03,Fournier 03,Kagawa 04}. Here
BEDT-TTF (or simply ET) denotes
bis(ethylenedithio)tetrathiafulvalene and X a monovalent anion.
For the latter substances, pressure studies revealed a first-order
metal-insulator transition line $T_{MI}(P)$ \cite{Ito 96,Lefebvre
00,Limelette 03,Fournier 03, Kagawa 04}, indicative of a
bandwidth-controlled Mott transition \cite{Kanoda 97a,Kino 95},
and suggest a second-order critical endpoint $(P_{0},T_{0})$
\cite{Lefebvre 00,Limelette 03,Fournier 03,Kagawa 04} with
remarkable properties. Particularly striking was the observation
of a pronounced softening of the $c_{22}$ elastic mode
\cite{Fournier 03}. Although acoustic and lattice anomalies are
expected \cite{Hassan 05, Merino 00} at the Mott transition in
\textit{response} to the softening of the electronic degrees of
freedom, the actual role of the lattice for the Mott transition in
real materials remains illusive. In addition, an unconventional
Mott criticality was proposed for the present organic salts
\cite{Kagawa 05} and attributed to their quasi-twodimensional
(quasi-2D) electronic character.

In this Letter we report, for the first time, the direct
observation of lattice anomalies at the Mott transition in a
$\kappa$-(ET)$_{2}$X organic conductor and explore, via a
sensitive thermodynamic probe, the criticality near
$(P_{0},T_{0})$.

For the thermal expansion measurements, a high-resolution
capacitive dilatometer (built after \cite{Pott 83}) was used,
enabling the detection of length changes $\Delta l$ $\geq$
10$^{-2}$\,\AA. Owing to the experimental difficulties posed by
accomplishing high-resolution dilatometric measurements under
variable pressure, use was made of the possibility of applying
chemical pressure. To this end, single crystals of
$\kappa$-(d8-ET)$_{2}$Cu[N(CN)$_{2}$]Br were synthesized with
deuterium atoms replacing the protons in the terminal ethylene
groups. These fully deuterated salts, referred to as d8-Br in the
following, are known to be situated very close to the MI
transition \cite{Kawamoto 97}. First, deuterated (98\%) ET
molecules were prepared according to \cite{Hartke 80, Mizuno 78}
using multiple recrystallization for the intermediate steps. Next,
single crystals were synthesized along an alternative preparation
route described recently for the protonated variant h8-Br
\cite{Strack 05}. The grade of deuteration was checked by infrared
reflection spectroscopy both on the deuterated ET material
\cite{Gaertner 91} as well as on the d8-Br single crystals
\cite{Griesshaber 00}, and found to be at least 98\%. For the
present study, crystals of two independently prepared batches were
used: crystal \#1 (batch A2907) and \#3 (A2995). The crystals have
the shape of flat distorted hexagons with dimensions of about 1
$\times$ 1 $\times$ 0.4 mm$^{3}$. The pressure exerted on the
crystal by the dilatometer ranges from 1 to 6 bar. A preliminary
account on a second crystal from batch A2907 was given in
\cite{Lang 06}. The resistivity was studied by employing a
standard four-terminal ac-technique. All measurements, unless
stated otherwise, were carried out after cooling through the glass
transition at $T_{g} \sim$ 77\,K with a very low rate of -3\,K/h
(thermal expansion) and -6\,K/h (resistivity) to rule out
cooling-rate dependent effects, see \cite{Mueller 02}.

\begin{figure}[floatfix]
\begin{center}
  \includegraphics[width=0.9\columnwidth]{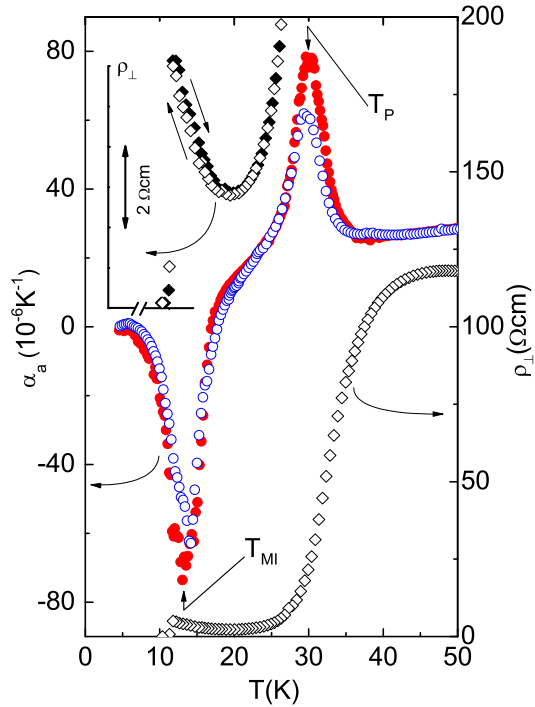}\\[-0.5cm]
   \caption{(Color online) In-plane $a$-axis expansivity, $\alpha_{a}$ (full circles),
   (left scale) and interlayer resistivity, $\rho_{\perp}$, (right scale) for
 $\kappa$-(d8-ET)$_{2}$Cu[N(CN)$_{2}$]Br crystal \#1 together with $\alpha_{a}$
 data for crystal \#3 (open circles). Upper inset: blow-up of
 low-$T$ $\rho_{\perp}(T)$ data on the same $T$ scale as used in the main panel.}
 \label{fig:1}
\end{center}
\end{figure}

\begin{figure}
\includegraphics[width=.90\columnwidth]{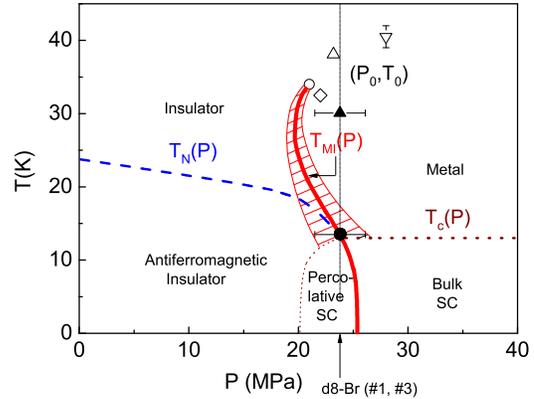}\\[-0.5cm]
\caption{(Color online) Phase diagram of $\kappa$-(ET)$_{2}$X
including the N\'{e}el transition at $T_{N}$ (dashed line) and the
superconducting (SC) transition at $T_{c}$ (dotted line) from
\cite{Lefebvre 00}. Thin solid lines, delimiting the hatched area,
denote positions of acoustic anomalies \cite{Fournier 03}. The
middle position (thick solid line) is used here as $T_{MI}(P)$
\cite {Note1}. Closed symbols refer to anomalies at $T_{MI}$ and
$T_{p} \simeq T_{0}$ of the d8-Br crystals \#1 and \#3 (same
positions on the scale of the figure), while open symbols denote
literature results for $(P_{0},T_{0})$: ($\diamond$)
\cite{Lefebvre 00}, ($\bigtriangledown$) \cite{Limelette 03},
($\bigcirc$) \cite{Fournier 03}, ($\bigtriangleup$) \cite{Kagawa
04}. Vertical line indicates the $T$-sweeps performed here.}
\label{fig:3}
\end{figure}

The interlayer resistivity $\rho_{\perp}$ for crystal \#1 is shown
in the lower part of Fig.\,\ref{fig:1}. Upon cooling,
$\rho_{\perp}$ passes over a maximum around 45\,K, then rapidly
drops and flattens around 30\,K. The resistivity remains metallic
down to about 20\,K, below which the slope sharply increases (cf.\
upper inset in Fig.\,\ref{fig:1}) indicating the transition into
an insulating state. A similar $\rho_{\perp}$ was found for
crystal \#3 and the crystal studied in \cite{Lang 06} including
the vanishing of $\rho_{\perp}$ below about 11.5\,K. A zero
resistivity accompanied by a tiny signature in the $\alpha_{i}$
data is consistent with percolative superconductivity in a minor
metallic phase coexisting with an antiferromagnetic/insulating
ground state for d8-Br \cite{Miyagawa 02}, cf.\ the phase diagram
in Fig.\,\ref{fig:3}.

\begin{figure}
\includegraphics[width=\columnwidth]{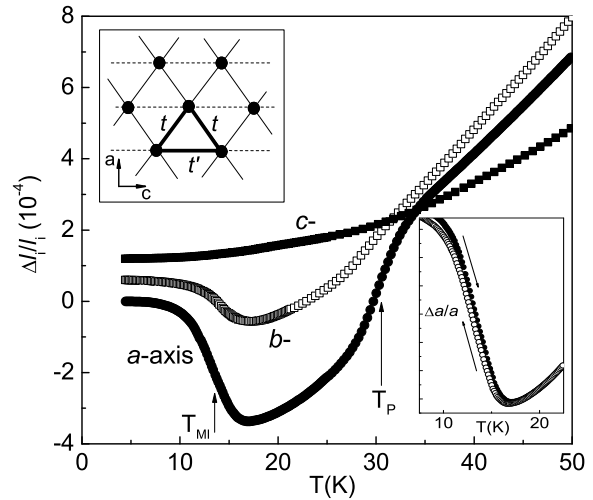}\\[-0.5cm]
\caption{Relative length changes for
$\kappa$-(d8-ET)$_{2}$Cu[N(CN)$_{2}$]Br crystal \#1 along the
in-plane $a$- and $c$-, and out-of-plane $b$-axis. The data have
been offset for clarity. Lower inset shows hysteretic behavior in
$\Delta a / a$ at $T_{MI}$ measured at very low sweeping rates of
$\pm$1.5\,K/h. Upper inset depicts the 2D triangular-lattice dimer
model with transfer integrals $t$ and $t'$.} \label{fig:2}
\end{figure}

The main features in the resistivity have their clear
correspondence in the coefficient of thermal expansion, $\alpha =
l^{-1}dl/dT$, also shown in Fig.\,\ref{fig:1} along the $a$-axis
of crystal \#1. The flattening of $\rho_{\perp}$ is accompanied by
a huge peak in $\alpha (T)$ centered at a temperature referred to
as $T_{p}$ in the following around 30\,K. As will be discussed
below, this effect can be assigned to a second-order phase
transition. Upon further cooling, $\alpha_{a}(T)$ reveals an even
bigger negative peak indicating yet another phase transition. The
accompanying change in $\rho_{\perp}$ from metallic to insulating
behavior suggests this peak to be due to the MI transition. This
is consistent with measurements under magnetic fields up to 10\,T
(not shown) leaving the peak position unaffected. A very similar
$\alpha _{a}(T)$ behavior is observed for \#3, although with
slightly reduced ($\sim 20\%$) peak anomalies and minor shifts in
$T_{p}$ and $T_{MI}$, cf.\ Fig.\,\ref{fig:1}. More insight into
the character of the transitions can be gained by looking at the
relative length changes $\Delta l_{i}(T)/l_{i}$ = $(l_{i}(T)$ -
$l_{i}(300\,K))/l_{i}(300\,K)$, ($i = a,b,c$) shown in
Fig.\,\ref{fig:2} for crystal \#1. The dominant effects occur
along the in-plane $a$-axis, i.e.\ parallel to the anion chains.
Here a pronounced s-shaped anomaly is revealed at $T_{p}$ which
lacks any sign of hysteresis upon cooling and warming -- generic
features of a second-order phase transition with strong
fluctuations. On further cooling through $T_{MI}$, the $a$-axis
shows a rapid increase of about $\Delta a/a = 3.5 \cdot 10^{-4}$
within a narrow temperature range, indicative of a slightly
broadened first-order transition. The observation of a small but
significant hysteresis of about 0.4\,K (cf.\ lower inset
Fig.\,\ref{fig:2}), which complies with the hysteresis in
$\rho_{\perp}(T)$ (upper inset Fig.\,\ref{fig:1}), confirms the
first-order character. The corresponding anomalies along the
$b$-axis are less strongly pronounced. Surprisingly, for the
second in-plane $c$-axis, anomalous behavior in $\Delta l /l$ can
be discerned neither at $T_{p}$ nor at $T_{MI}$. The same
anisotropy was found for the second crystal of batch A2907 studied
in \cite{Lang 06} (not shown), on which all three uniaxial
expansion coefficients had been determined.

Figure \ref{fig:2} reveals that the anomalies at $T_{p}$ and
$T_{MI}$ are correlated in size, albeit with reversed sign,
suggesting that they are intimately related to each other. In
addition, the data disclose a striking in-plane anisotropy. Given
the quasi-2D electronic structure as shown in the upper inset of
Fig.\,\ref{fig:2}, characterized by dimers on an anisotropic
triangular lattice \cite{Kino 95}, the latter is a very remarkable
and unexpected result: The dominant response in the $a$-axis,
along which no direct dimer-dimer overlap exists, means that the
diagonal electronic interactions along the $c \pm a$ directions,
$t$, have to be involved in this process. Since these interactions
have a strong component also along the $c$-axis, which is likely
to be even softer than the anion-chain $a$-axis \cite{Cu(NCS)}, a
significant $c$-axis response would be expected at $T_{MI}$. A
zero-effect along the $c$-axis is even more amazing as there is a
relatively strong direct dimer-dimer interaction $t'$ along this
axis, cf.\ upper inset of Fig.\,\ref{fig:2}. Thus, to account for
a zero $c$-axis response within a 2D electronic model would imply
an accidental cancellation of counteracting effects associated
with $t$ and $t'$, which seems very unlikely. Furthermore, it is
not obvious how these in-plane interactions may cause the
comparatively strong effect in the interlayer $b$-axis, along
which the lattice is expected to be even more stiff
\cite{Cu(NCS)}. These observations suggest that a coupling of the
$\pi$-electrons to other degrees of freedom has to be taken into
account to understand the MI transition here.

Before discussing further implications of our observations, the
MI-transition temperature $T_{MI}$ is determined. As
Fig.\,\ref{fig:2} demonstrates, the transition is not very sharp
but rather spans a range of several Kelvin -- an effect which is
very similar for both crystals studied here and the one explored
in \cite{Lang 06}. A broadening of signatures in $T$-dependent
measurements, as opposed to isothermal pressure sweeps, would be
naturally expected given the steepness of $T_{MI}(P)$, cf.\
Fig.\,\ref{fig:3}. However, the width of about 5.6\,K of the
$\Delta l_{a}/l_{a}$ jump (10-90\%), which transforms into a
pressure interval of about 2\,MPa employing a slope $dT_{MI}/dP =
-2.7 \pm 0.1 $\,K/MPa around 14\,K (cf.\ Fig.\,\ref{fig:3}), is
even smaller than the transition range seen in acoustic
measurements as a function of pressure \cite{Fournier 03} (cf.\
hatched area in Fig.\,\ref{fig:3}), but is comparable with a width
of about 1.4\,MPa as read off the resistivity data in \cite{Kagawa
04}. These smearing effects have been attributed to a region of
coexistence between insulating and metallic phases \cite{Limelette
03}, as indeed observed via real-space imaging \cite{Sasaki 05}.
For lack of a well-founded procedure to treat the broadened
transitions, the position of the $\alpha (T)$ minimum is chosen as
the thermodynamic transition temperature. Employing literature
results on $T_{MI}(P)$ \cite{Note1}, the so-derived values of
$T_{MI}$ = $(13.5 \pm 0.8)$\,K, (\#1) and $(14.1 \pm 0.8)$\,K,
(\#3) can be used to pinpoint the position of the present d8-Br
crystals on the pressure axis in Fig.\,\ref{fig:3}. Within the
uncertainties implied in this procedure, the crystals are located
very close to the critical pressure $P_{0}$ as determined by the
various pressure studies \cite{Lefebvre 00, Limelette 03, Fournier
03, Kagawa 04}. The significance of this finding is twofold.
First, it demonstrates that the anomaly at $T_{p}$ reflects the
lattice response at $(P_{0},T_{0})$. Second, as this point is part
of the $T_{MI}$ line, it provides a natural explanation for the
intimate interrelation of the anomalies at $T_{p}$ and $T_{MI}$
inferred from Fig.\,\ref{fig:2}.

\begin{figure}[floatfix]
  \includegraphics[width=\columnwidth]{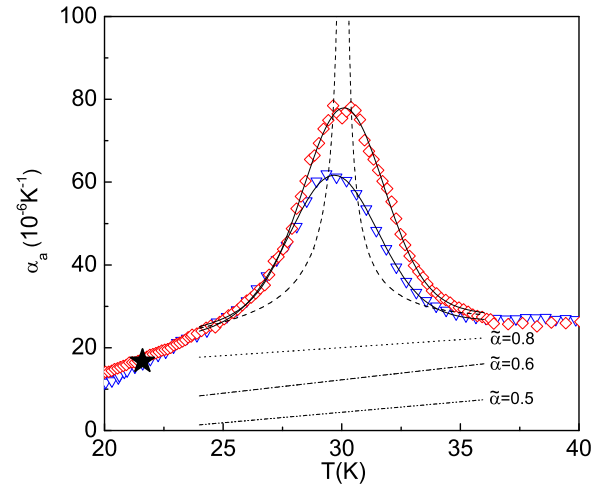}\\[-0.5cm]
  \caption{(Color online) Expansivity along the $a$-axis for d8-Br crystals
  \#1 ($\diamond$) and \#3 ($\bigtriangledown$) near $T_{0}$. Solid (dashed) lines are fits as described
  in the text for $\tilde{\alpha}$ = 0.8 with (without for \#1) a Gaussian distribution of
  $T_{0}$. Straight lines show background contributions implied in fits
  for $\tilde{\alpha}$ values given in the figure. Star marks universal background point discussed in the text.} \label{fig:4}
\end{figure}

The huge anomaly at $T_{p} \simeq T_{0}$, exceeding the background
by a factor 3-4, enables us to explore the criticality at
$(P_{0},T_{0})$ with extraordinarily high sensitivity. To this
end, the phase transition anomaly in $\alpha_{a}(T)$, shown for
the crystals \#1 and \#3 in Fig.\,\ref{fig:4} on expanded scales,
is analyzed in terms of a power-law behavior in the variable $t =
(T - T_{0})/T_{0}$. This approach is based on the proportionality
of $\alpha (T)$ to the specific heat, $C(T)$, implying that the
same scaling laws apply at $T_{0}$, as verified by various groups,
see, e.g.\,\cite{Pasler 98, Souza 05}.

The data sets in Fig.\,\ref{fig:4} reveal a steep increase in the
slope of $\alpha(T)$ on the outer flanks of the maximum. Closer to
the center of the peak, however, the slope is reduced giving rise
to a rounded maximum. Such broadening effects are generally
encountered in the immediate vicinity of the transition and
attributed to sample inhomogeneities. The rounding over a
considerable temperature range here demands particular attention.
For the description of the data in the range 24 - 36\,K, the
function $\alpha =
\frac{A^{\pm}}{\tilde{\alpha}}|t|^{-\tilde{\alpha}} + B + E \cdot
T$ was used. This function contains the singular contribution with
the amplitudes $A^{+}$ and $A^{-}$ for $t
> 0$ and $t < 0$, respectively, and a linear term. The
latter comes primarily from the phonons but can also include a
small non-singular electronic contribution. The smearing of the
transition is accounted for by a Gaussian distribution for
$T_{0}$, $G(\bar{T_{0}}, T_{0}, \delta T_{0})$, centered at
$\bar{T_{0}}$ with a width $\pm \delta T_{0}$. Applying the
function
$\int[\frac{A^{\pm}}{\tilde{\alpha}}|(T-T_{0})/T_{0}|^{-\tilde{\alpha}}
+ B + E \cdot T] \cdot G(\bar{T_{0}}, T_{0}, \delta T_{0})\cdot
dT_{0}$, the data sets of crystal \#1 and \#3 were fitted
simultaneously using the same exponent $\tilde{\alpha}$, the same
ratio $A^{+}/A^{-}$, and an identical background for both
crystals. A constraint for the background contribution can be
derived by comparing the data in Fig.\,\ref{fig:4}, with those of
the d8-Br in \cite{Lang 06} and h8-Br in \cite{Mueller 02}. All
data sets intersect at a single point $T \simeq$ 21\,K, $\alpha
\simeq$ 16 $\cdot$ 10$^{-6}$K$^{-1}$, irrespective of the presence
and size of the critical contribution at $T_{0}$, indicating that
this point reflects the pure background. Thus a meaningful
background should extrapolate to this universal point. A good fit
to both data sets, also satisfying this background constraint, is
obtained for $\tilde{\alpha}$ = 0.8, $A^{+}/A^{-}$ = 0.79, and
$\bar{T_{0}}$ = 30.1\,K, $\delta T_{0}$ = 1.59\,K for \#1 and
$\bar{T_{0}}$ = 29.6\,K, $\delta T_{0}$ = 1.74\,K for \#3, cf.\
Fig.\,\ref{fig:4}. We stress that $\tilde{\alpha}$ values in the
range 0.65 - 0.95, with small changes in the other parameters
accordingly, result in fits of similar quality and still comply
with the background constraint. In contrast, the residual of the
fit increases substantially upon decreasing $\tilde{\alpha}$ to
well below 0.65. This is accompanied by a suppression of the
background to even negative values for $\tilde{\alpha} <$ 0.5,
clearly incompatible with the background constraint, see, e.g.,
the background implied in the fits for $\tilde{\alpha}$ = 0.6 and
0.5 in Fig.\,\ref{fig:4}. As clearly indicated by these
simultaneous fits, and confirmed by independent fits to the
individual data sets for crystals \#1 and \#3, a large positive
$\tilde{\alpha}$ value is the only possible, physically meaningful
description of the expansivity data.

The critical exponent derived here of $\tilde{\alpha} \simeq$ 0.8
$\pm$ 0.15 is much larger than those of known universality classes
with -0.12 $\leq \tilde{\alpha} \leq$ 0.14  and the mean-field
value $\tilde{\alpha}$ = 0 observed at the Mott critical endpoint
of Cr-doped V$_{2}$O$_{3}$ \cite{Limelette 03a}. In particular, it
greatly conflicts with the criticality reported in \cite{Kagawa
05} for pressurized X = Cu[N(CN)$_{2}$]Cl. Employing the exponent
identity $\tilde{\alpha} + 2\beta + \gamma = 2$ \cite{Kadanoff 67}
the exponents found there of $(\delta, \beta, \gamma) \approx (2,
1, 1)$ give $\tilde{\alpha}$ = -1. The reason for this discrepancy
is unclear but might be related to the significant broadening
effects \cite{Note2}, which have not been included in the analysis
in \cite{Kagawa 05}. The exponent found here, however, is rather
close to $\tilde{\alpha}$ = 0.5 expected for a tricritical point
\cite{Huang 87}. Such a scenario would imply a symmetry breaking
associated with $T_{MI}$ for which no evidence has yet been
supplied. Interestingly, an even larger exponent $\tilde{\alpha}$
= 0.93 was reported for La$_{0.7}$Ca$_{0.3}$MnO$_{3}$ \cite{Souza
05}, also characterized by a strong electron-phonon coupling,
showing a similar $\alpha(T)$ anomaly as the one observed here.

In summary, high-resolution dilatometry on deuterated
$\kappa$-(ET)$_{2}$Cu[N(CN)$_{2}$]Br crystals reveals
discontinuous changes of the lattice parameters at the Mott
transition. The data disclose a striking anisotropy unlikely to be
captured by a 2D purely electronic model. An analysis of the huge
thermal expansion anomaly at the end-point of the first-order
$T_{MI}$ line yields a critical exponent $\tilde{\alpha } \simeq$
0.8 $\pm$ 0.15, markedly different from the criticality derived
from transport measurements \cite{Kagawa 05}. The unusually large
$\tilde{\alpha }$ value together with the anomalous anisotropy of
the lattice effects at $T_{MI}$ suggest an intricate role of the
lattice in the Mott transition for the present materials.

\begin{acknowledgments}
M. de Souza acknowledges financial support from the Brazilian
Research Foundation CAPES and the DAAD.
\end{acknowledgments}


\begin{thebibliography}{Buravov 271}
\bibitem{Cu(NCS)}For the related X = Cu(NCS)$_{2}$ salt, the uniaxial
compressibilities $k_{i}$ are strongly anisotropic \cite{Chasseau
91} with $k_{1}:k_{2}:k_{3}=1:0.53:0.17$, and $i$ = 1 and 2 the
in-plane axis perpendicular and parallel to the anion chains,
respectively, and $i$ = 3 along the long axis of the ET molecules.
\end{thebibliography}
\end{document}